\documentclass[preprint, superscriptaddress, preprintnumbers,amsmath,amssymb]{revtex4}

\allowdisplaybreaks
\allowdisplaybreaks[4]
\usepackage{CJK}
\usepackage{graphicx}% Include figure files
\usepackage{dcolumn}% Align table columns on decimal point
\usepackage{bm}% bold math
\usepackage{makecell}
\begin{document}

\begin{CJK}{GBK}{song}

\title{Chiral geometry and rotational structure for $^{130}$Cs in the projected shell model}

\author{F. Q. Chen}
\affiliation{State Key Laboratory of Nuclear Physics and Technology, School of Physics, Peking University,
Beijing 100871, People's Republic of China}
\author{J. Meng}
 \email{mengj@pku.edu.cn}
\affiliation{State Key Laboratory of Nuclear Physics and Technology, School of Physics, Peking University,
Beijing 100871, People's Republic of China}
\affiliation{Yukawa Institute for Theoretical Physics, Kyoto University, Kyoto 606-8502, Japan}
\affiliation{Department of Physics, University of Stellenbosch, Stellenbosch, South Africa}
\author{S. Q. Zhang}
 \email{sqzhang@pku.edu.cn}
\affiliation{State Key Laboratory of Nuclear Physics and Technology, School of Physics, Peking University,
Beijing 100871, People's Republic of China}

\date{\today}

\begin{abstract}
The projected shell model with configuration mixing for nuclear chirality is developed and applied to the observed rotational bands in the chiral nucleus $^{130}$Cs. For the chiral bands, the energy spectra and electromagnetic transition probabilities are well reproduced. The chiral geometry illustrated in the $K~plot$ and the $azithumal~plot$ is confirmed to be stable against the configuration mixing. The other rotational bands are also described in the same framework.
\end{abstract}

\maketitle

%\subsection{INTRODUCTION}

Spontaneous chiral symmetry breaking in atomic nuclei has attracted intensive theoretical and experimental studies since its first prediction by Frauendorf and Meng \cite{TAC1997} in 1997. It takes place when the angular momentum of a triaxial nucleus has non-vanishing components on all of the three intrinsic principle axes. These three components, contributed by the valence proton(s), the valence neutron(s), and the rotating core, respectively, can form a left- or a right- handed configuration.

The picture of chiral geometry is described in the intrinsic frame, and it is manifested in the laboratory system by the observation of the chiral doublet bands, which are a pair of near-degenerate $\Delta I=1$ bands with the same parity. Intensive efforts have been devoted to the search of chiral doublet bands in various mass regions. So far around 50 pairs of chiral doublets have been discovered in $A\sim80$ \cite{Br80exp,Br78exp}, $100$ \cite{Rh104exp,Ag106exp1,Rh102exp,Ag106exp3,Ag106exp2,Rh103exp}, $130$ \cite{Starosta2001,Nd135exp,Pr134exp1,Pr134exp2,Cs128-2006,Nd135exp2,Ce133exp} and $190$ \cite{Ir188exp,Tl198exp} mass regions, see Refs. \cite{Meng2010JPG,Meng2014IJMPE,Meng2016PS} for reviews. The chiral doublet bands observed experimentally up to now has been compiled very recently in Ref. \cite{Xiong2018}. Theoretical studies of the chiral doublet bands have been carried out by the particle rotor model (PRM) \cite{TAC1997,Pengjing2003,Koike2004,ZSQ2007,QB2009PLB}, the tilted axis cranking model (TAC) \cite{TAC1997,Dimitrov2000TAC,Madokoro2000,SkyrmeTAC1,SkyrmeTAC2,ZPW2017}, the TAC plus random phase approximation \cite{RPA2011}, the collective Hamiltonian \cite{CQB2013,CQB2016}, the interacting boson-fermion-fermion model (IBFFM) \cite{Pr134exp1,Tonev2007,Brant2008}, and the generalized coherent state model \cite{CoherentJPG2016}.

The angular momentum projection (AMP) approach, which restores the broken rotational symmetry in the mean field wavefunctions, is a promising tool for the microscopic description to the nuclear system. The AMP approach could be based on mean field descriptions from Nilsson+BCS \cite{Hara1995} to various density functionals \cite{Sabbey2007,Rodriguez2010,Rodriguez2015,Niksic2007,YJM2010}, and have been applied to various problems as summarized in a recent review \cite{Sunyang2016}. As one of the implementations of the AMP approach, the projected shell model \cite{Hara1995} has been used in attempts to understand the chiral doublet bands \cite{Bhat2012,Bhat2014}. The observed energy spectra and electromagnetic transitions for the doublet bands are well reproduced in these calculations. However, it was found difficult to give an illustration for the underlying chiral geometry. The difficulty lies in the fact that the angular momentum geometry is defined in the intrinsic frame, while the angular momentum projected wave functions are written in the laboratory frame. This difficulty is overcomed recently by the introduction of the $K~plot$ and the $azimuthal~plot$ \cite{CFQ2017}. The $K~plot$ and the $azimuthal~plot$ provide probability distributions of the orientation related quantities in the intrinsic frame, as demonstrated by the chiral doublet bands in $^{128}$Cs \cite{CFQ2017}. In the angular momentum projected framework, the chiral geometry can also be illustrated by the root-mean-square values of the angular momentum components in the intrinsic frame \cite{Shimizu2018}.
%The reason lies in the fact that operator(s) representing the intrinsic orientation of the angular momentum could not be defined in the laboratory system, in which the angular momentum projected wave functions are written. To examine the chiral geometry in the AMP framework, the best thing one can do is to find quantities in the angular momentum projected wave functions that relates to the intrinsic orientation of the angular momentum. Even if they are not eigen values of any physical operators, the distributions of these quantities may be obtained from the wave functions, by which the angular momentum geometry can be revealed.

%The above idea has been realized very recently in Ref. \cite{CFQ2017}, in which the chiral doublet bands in $^{128}$Cs were studied by the AMP approach, and the chiral geometry was illustrated by the $K$-Plot and the Azimuthal-Plot. The $K$-Plot gives the distributions of the components of the angular momentum on the three intrinsic axes, and the Azimuthal-Plot gives the distributions of the tilted angles, i.e., the polar angle and the azimuthal angle of the angular momentum in the intrinsic frame. With these plots the chiral symmetry breaking in $^{128}$Cs is demonstrated.

%The yrast band is obtained by blocking
%the lowest ¦Ðh11/2 and the fourth ¦Íh11/2 orbitals,

The calculation in Ref. \cite{CFQ2017} is done by blocking
the lowest $\pi h_{11/2}$ and the fourth $\nu h_{11/2}$ orbitals,
%based on a fixed intrinsic configuration, therefore it is a simplified description.
and the configuration mixing is neglected. For a better description to the nuclear system, the theoretical framework in Ref. \cite{CFQ2017} needs to be generalized to include various configurations, i.e., to the projected shell model. In order to have a better understanding of the chiral doublet bands, it is necessary to confirm whether the chiral geometry is stable against configuration mixing. Another advantage is that the projected shell model can provide a simultaneous description for all rotational bands in one nucleus on the same foot.

In this work the projected shell model with configuration mixing for nuclear chirality is developed and applied to $^{130}$Cs \cite{Starosta2001}. The chiral geometry for the chiral doublet bands is illustrated in terms of the $K~plot$ and the $azimuthal~plot$, in a similar way as Ref. \cite{CFQ2017}.  The effect of the configuration mixing on the chiral geometry and the description of the other rotational bands are discussed.

%\subsection{THEORETICAL FRAMEWORK}

The frame work of the projected shell model is based on the standard pairing plus quadrupole Hamiltonian \cite{ManyBody},
\begin{equation}\label{Hamiltonian}
\hat{H}=\hat{H}_0-\frac{\chi}{2}\sum_{\mu}\hat{Q}^+_{\mu}\hat{Q}_{\mu}-G_{M}\hat{P}^+\hat{P}-G_{Q}\sum_{\mu}\hat{P}^+_{\mu}\hat{P}_{\mu},
\end{equation}
which includes a spherical single-particle part and the separable two body interactions, i.e., the quadrupole-quadrupole interaction, the monopole pairing, and the quadrupole pairing. The intrinsic ground state of an odd-odd nucleus can be denoted as $|\Phi_{\nu_0\pi_0}\rangle$, where $\nu_0$ and $\pi_0$ are single-particle orbitals blocked. The state $|\Phi_{\nu_0\pi_0}\rangle$ is determined by the variation principle with the constraints on the quadrupole momentes and the average particle numbers:
\begin{equation}\label{variation}
\delta\langle\Phi_{\nu_0\pi_0}|\hat{H}-\lambda_{q_0}\hat{Q}_0-\lambda_{q_2}\hat{Q}_2-\lambda_N\hat{N}-\lambda_Z\hat{Z}|\Phi_{\nu_0\pi_0}\rangle=0,
\end{equation}
and can be written as a two-quasiparticle state on top of a quasiparticle vacuum of the even-even core $|\Phi_0\rangle$:
\begin{equation}\label{intrinsicgs}
|\Phi_{\nu_0\pi_0}\rangle=\beta^+_{\nu_0}\beta^+_{\pi_0}|\Phi_0\rangle.
\end{equation}
By the variation in Eq. (\ref{variation}), the state $|\Phi_{\nu_0\pi_0}\rangle$ and the vacuum $|\Phi_0\rangle$ are obtained, as well as corresponding quasiparticle operators $\{\beta^+_\nu,\beta^+_\pi\}$, with $\nu$ and $\pi$ the single-particle orbitals. Various two-quasiparticle states can be constructed as $\{|\Phi_\kappa\rangle\}=\{\beta^+_\nu\beta^+_\pi|\Phi_0\rangle\}$, in which $\kappa$ specifies different two-quasiparticle configurations. The effect of the configuration mixing, which was neglected in Ref. \cite{CFQ2017}, can be then taken into account.

The two-quasiparticle states $|\Phi_{N,Z,\kappa}\rangle$ with good particle number $N$ and $Z$ can be projected from $|\Phi_\kappa\rangle$:
\begin{equation}\label{NZprojection}
|\Phi_{N,Z,\kappa}\rangle\equiv\hat{P}^N\hat{P}^Z|\Phi_\kappa\rangle.
\end{equation}
The symmetry restored basis is constructed by the angular momentum projection:
\begin{equation}\label{AMPbasis}
\{\hat{P}^I_{MK}|\Phi_{N,Z,\kappa}\rangle\}.
\end{equation}
%In Eq. (\ref{AMPbasis}), the $K$ values allowed for each two-quasiparticle state $|\Phi_\kappa\rangle=\beta^+_\nu\beta^+_\pi|\Phi_0\rangle$ are determined by the orbitals $\nu$ and $\pi$ \cite{Hara1995}. The quasiparticle operators $\beta^+_\tau$ ($\tau=\nu,\pi$) satisfy
%\begin{equation}\label{zpisymmetry}
%e^{-i\pi\hat{J}_z}\beta^+_{\tau}e^{i\pi\hat{J}_z}=\pm i\beta^+_\tau.
%\end{equation}
%If both $\beta^+_{\nu}$ and $\beta^+_{\pi}$ in $|\Phi_\kappa\rangle$ satisfy Eq. (\ref{zpisymmetry}) with the same sign, $K$ takes only odd values. If $\beta^+_{\nu}$ and $\beta^+_{\pi}$ satisfy Eq. (\ref{zpisymmetry}) with opposite signs, $K$ takes only even values.

The diagonalization of the Hamiltonian in the projected basis (\ref{AMPbasis}) leads to the Hill-Wheeler equation:
\begin{equation}\label{HillWheeler}
\sum_{K'\kappa'}(\langle\Phi_{N,Z,\kappa}|\hat{H}\hat{P}^I_{KK'}|\Phi_{N,Z,\kappa'}\rangle-E^{I\sigma}\langle\Phi_{N,Z,\kappa}|\hat{P}^I_{KK'}|\Phi_{N,Z,\kappa'}\rangle) f^{I\sigma}_{KK'}=0,
\end{equation}
in which $\sigma$ specifies different eigen states of the same spin $I$. By solving Eq. (\ref{HillWheeler}) the eigenenergies $E^{I\sigma}$ and the wave functions
\begin{equation}\label{wavefunction}
|\Psi^{\sigma}_{IM}\rangle=\sum_{K\kappa}f^{I\sigma}_{K\kappa}\hat{P}^I_{MK}|\Phi_{N,Z,\kappa}\rangle
\end{equation}
are obtained. With the wave functions (\ref{wavefunction}) the electromagnetic transitions and other physical quantities can be calculated.

In order to examine the configuration mixing, it is necessary to know the weights of different two-quasiparticle configurations $|\Phi_\kappa\rangle$ in the wave function (\ref{wavefunction}). These weights could be obtained by resorting the concept of collective wave functions in the generator coordinate method (GCM), and  regarding $\{K,\kappa\}$ in Eq. (\ref{wavefunction}) as generator coordinates. The corresponding generating functions are the projected basis in Eq. (\ref{AMPbasis}), and the corresponding norm matrix elements write
\begin{equation}\label{normKkappa}
\mathcal{N}_{I}(K,\kappa;K',\kappa')\equiv\langle\Phi_{N,Z,\kappa}|\hat{P}^I_{KK'}|\Phi_{N,Z,\kappa'}\rangle.
\end{equation}
The collective wave functions $g^{I\sigma}(K,\kappa)$ can be obtained by the square root of the norm matrix (\ref{normKkappa}) and the function $f^{I\sigma}_{K\kappa}$:
\begin{equation}\label{collectiveKkappa}
g^{I\sigma}(K,\kappa)=\sum_{K',\kappa'}\mathcal{N}^{1/2}_{I}(K,\kappa;K',\kappa')f^{I\sigma}_{K'\kappa'},
\end{equation}
and they are proved to satisfy the normalization and orthogonal relation
\begin{equation}\label{Kkappanormalize}
\sum_{K\kappa}g^{I\sigma_1*}(K,\kappa)g^{I\sigma_2}(K,\kappa)=\delta_{\sigma_1\sigma_2}.
\end{equation}
In the GCM, the collective wave functions are understood as probability amplitudes of the generator coordinates. Therefore, the weight of the configuration $|\Phi_\kappa\rangle$ in the state $|\Psi^\sigma_{IM}\rangle$ could be written as:
\begin{equation}\label{cfgweights}
W_\kappa=\sum_K|g^{I\sigma}(K,\kappa)|^2.
\end{equation}
The collective wave functions (\ref{collectiveKkappa}) are also used in the calculations of the $K~plot$ and the $azimuthal~plot$, which have been introduced in Ref. \cite{CFQ2017}.

In the following, the five rotational bands observed in $^{130}$Cs \cite{Cs130NPA,Cs130EPJA,Cs130transitions}, including the chiral doublet bands, are investigated by the present approach. The parameters in the Hamiltonian (\ref{Hamiltonian}) are taken from Ref. \cite{GZC2006}. The quadrupole deformation parameters ($\beta,\gamma$) are constrained to be ($0.20, 30^\circ$). This choice of $(\beta,\gamma)$ is the same as that adopted in the calculation for $^{128}$Cs in Ref. \cite{CFQ2017}. It agrees reasonably with the deformations ($0.19, 39^\circ$) used in the projected shell model calculation in Ref. \cite{GZC2006}.

Different from the calculation in Ref. \cite{CFQ2017}, where the lowest $\pi h_{11/2}$ and the forth $\nu h_{11/2}$ orbitals are blocked, here the two-quasiparticle configurations $\beta^+_\nu\beta^+_\pi|\Phi_0\rangle$ are constructed using orbitals $\nu$ and $\pi$ from the $N=4$ and $N=5$ major shells. The quasiparticle energy cutoff $E_\nu+E_\pi\le3.5\text{MeV}$ is adopted for the two-quasiparticle configurations, in which $E_\nu$ and $E_\pi$ are the quasiparticle energies of the orbitals $\nu$ and $\pi$, respectively. There are 60 two-quasiparticle configurations with positive parity and 68 with negative parity taken into account in the configuration space. Half of them can be projected onto even $K$ values and the other half can be projected onto odd $K$ values \cite{Sheikh2008}. Therefore the dimension of the projected basis (\ref{AMPbasis}) for spin $I$ is $30(2I+1)$ for the positive parity subspace and $34(2I+1)$ for the negative parity subspace.

%\subsection{RESULTS AND DISCUSSION}

%\subsubsection{Spectra and electromagnetic transitions}
\begin{figure}[!h]
  \begin{center}
    \includegraphics[width=10 cm]{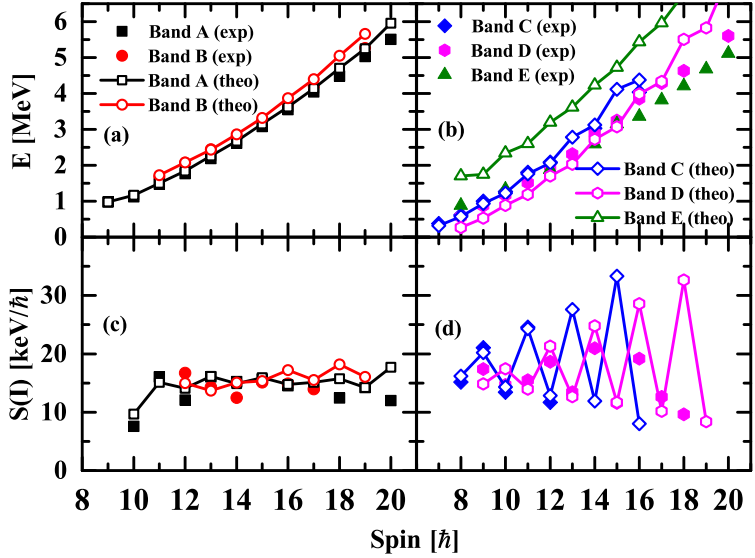}
    \caption{(Color online) (Upper panels) Experimental and calculated rotational bands in $^{130}$Cs. Panel (a) shows bands with positive parity while panel (b) shows those with negative parity. Experimental data are taken from Refs. \cite{Cs130NPA,Cs130EPJA,Cs130transitions}. The calculated energy levels are shifted by taking the level $9^+$ in band A as a reference. (Lower panels) Experimental and calculated values of $S(I)=[E(I)-E(I-1)]/2I$ for bands A, B (c) and bands C, D (d).}\label{spectra}
  \end{center}
\end{figure}
The calculated energy spectra of the rotational bands (denoted as bands A-E) in $^{130}$Cs are shown in Figs. \ref{spectra} (a) and (b), in comparison with the available data \cite{Cs130NPA,Cs130EPJA,Cs130transitions}.

As shown in Fig. \ref{spectra} (a), the calculated yrast and yrare bands with positive parity agree well with the experimental chiral doublet bands, band A and band B, including the near degeneracy between the partner bands. Note that the configuration mixing, which is absent in the calculation in Ref. \cite{CFQ2017}, is taken into account in the present calculation. The present results demonstrate that the near degeneracy between the chiral doublet bands is persistent even with the configuration mixing.

The calculated yrast and yrare bands with negative parity agree reasonably with the experimental bands D and C, as shown in Fig. \ref{spectra} (b). The levels with $I\ge18~\hbar$ in band D are overestimated, which might be due to a crossing with four-quasiparticle configurations and is beyond the present two-quasiparticle configuration space.

The calculated lowest band with the dominating configuration $\nu g_{7/2} \pi h_{11/2}$ suggested in Ref. \cite{Cs130EPJA} lies around 1 MeV above the experimental band E. The dominating configuration(s) will be explained in the following. One may notice that the moment of inertia for band E is larger than those for bands C and D, which suggests a larger deformation for band E. In this sense, band E is beyond the present scope. Therefore we exclude band E from the following discussions.

\begin{figure}[!h]
  \begin{center}
    \includegraphics[width=10 cm]{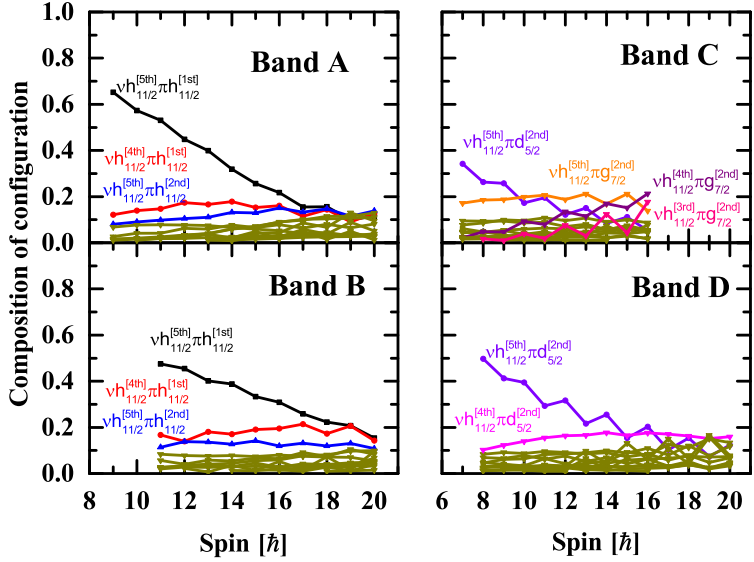}
    \caption{(Color online) Composition of configuration for bands A-D. For each band, contributions of the dominating configurations as functions of spin are presented by colored lines with their corresponding configuration labeled by the same color. Contributions of other configurations (if larger than $1\%$) are presented by dark yellow lines.}\label{configurations-ABCD}
  \end{center}
\end{figure}

The composition of each state can be calculated from Eq. (\ref{cfgweights}), and then the dominating configuration(s) can be recognized. The dominating configurations for bands A-D at the band head are given in Table \ref{table1}, together with those suggested in the previous studies. The dominating configurations found by the present calculation coincide with the previous assignments, except band C. The dominating configuration found for band C is $\nu h^{\text{[5th]}}_{11/2}\pi d^{\text{[2nd]}}_{5/2}$ while the previously assigned one is $\nu h_{11/2}\pi g_{7/2}$. This difference is due to the strong mixing between the $g_{7/2}$ and the $d_{5/2}$ orbitals as shown in Fig. \ref{configurations-ABCD}.

In Fig. \ref{configurations-ABCD}, the compositions of configurations for bands A-D are shown as functions of spin. For bands A and B, the configuration $\nu h_{11/2}^{[\text{5th}]}\pi h_{11/2}^{[\text{1st}]}$ is dominate until $I\ge16~\hbar$, after which strong configuration mixing occurs. For band C, the configurations $\nu h_{11/2}^{[\text{5th}]}\pi d_{5/2}^{[\text{2nd}]}$ and $\nu h^{[\text{5th}]}_{11/2}\pi g^{[\text{2nd}]}_{7/2}$ strongly compete with each other. The configuration $\nu h_{11/2}^{[\text{5th}]}\pi d_{5/2}^{[\text{2nd}]}$ wins at the band head, but the configuration $\nu h^{[\text{5th}]}_{11/2}\pi g^{[\text{2nd}]}_{7/2}$ takes over at $I=8~\hbar$. For band D, the configuration $\nu h_{11/2}^{[\text{5th}]}\pi d_{5/2}^{[\text{2nd}]}$ is dominant until $I=13~\hbar$, after which strong configuration mixing occurs.
%In all of the four panels of Fig. \ref{configurations-ABCD}, one may note the trend that quasiparticle orbitals at the middle of a $j$-shell become favoured at high spins. This is a reflection of rotational alignment, as the quasiparticles at the middle of a $j$-shell have angular momenta along the intermediate axis ($i$-axis), aligning with the orientation of the angular momentum of collective rotation.

The amplitudes of signature splitting of the bands A-D, reflected by the quantity $S(I)=[E(I)-E(I-1)]/2I$, are shown in Figs. \ref{spectra}(c) and (d). For bands A and B, the angular momentum of the valence neutron orientates along the long axis ($l$-axis), and that of the valence proton orientates along the short axis ($s$-axis). Both of them are prependicular approximately to the angular momentum of the collective rotation. The values of $S(I)$ stay almost independent of spin, as is seen in Fig. \ref{spectra}(c). In fact, the spin independence of $S(I)$ has been recognized as one of the criteria for the chiral doublet bands \cite{Rh104exp,WSY2007}. For bands C and D, the strong configuration mixing leads to the increasing signature splitting with spin \cite{ZPA1996} as shown in Fig. \ref{spectra}(d). Moreover, it is noted in Fig. \ref{spectra} that the amplitudes of $S(I)$ increase with spin, which is due to the increasing configuration mixing.

\begin{table}
\caption{The dominant configurations for bands A-D at the band head in the present calculation, in comparison with those suggested in the previous studies. The ordinal numbers in the second column denote specific single-particle orbitals. For example, the notation $\pi h_{11/2}^{[\text{1st}]}$ represents the first (lowest) single-particle orbital in the $\pi h_{11/2}$ subshell.}\label{table1}

\centering
\begin{tabular}{cccc}
  \hline
  Bands & \makecell{Present \\ configuration} & \makecell{Previous \\ configuration} & References \\\hline
  A, B & $\nu h_{11/2}^{[\text{5th}]}\pi h_{11/2}^{[\text{1st}]}$ & $\nu h_{11/2}\pi h_{11/2}$ & \cite{Starosta2001,Cs130transitions,Cs130EPJA} \\
  C & $\nu h_{11/2}^{[\text{5th}]}\pi d_{5/2}^{[\text{2nd}]}$ & $\nu h_{11/2}\pi g_{7/2}$ & \cite{Cs130NPA,Cs130EPJA} \\
  D & $\nu h_{11/2}^{[\text{5th}]}\pi d_{5/2}^{[\text{2nd}]}$ & $\nu h_{11/2}\pi d_{5/2}$ & \cite{Cs130NPA,Cs130EPJA} \\
%  E & $\nu g_{7/2}^{[\text{4th}]}\pi h_{11/2}^{[\text{1st}]}$ & $\nu g_{7/2}\pi h_{11/2}$ & \cite{Cs130EPJA} \\
  \hline
\end{tabular}
\end{table}

In the following we focus our discussion on the properties of the chiral doublet bands A and B.

\begin{figure}[!h]
  \begin{center}
    \includegraphics[width=8 cm]{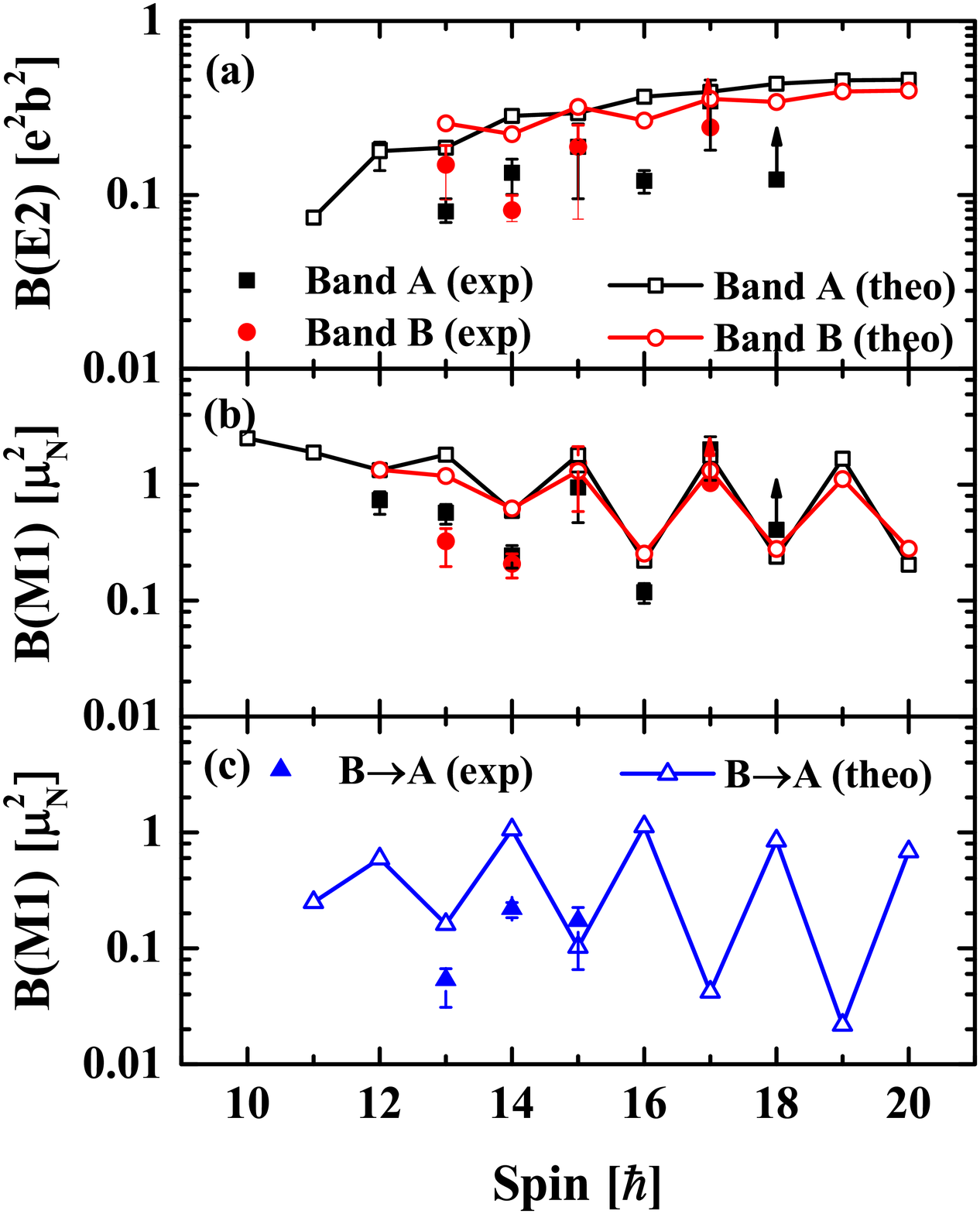}
    \caption{(Color online) (a) The intraband $B(E2)$, (b) the intraband $B(M1)$, and (c) the interband $B(M1)$ for the chiral doublet bands in $^{130}$Cs calculated by the projected shell model, compared with the experimental data taken from Ref. \cite{Cs130-ZhuLH}. }\label{transitions2}
  \end{center}
\end{figure}

The calculated intraband $B(E2)$ and $B(M1)$, and the interband $B(M1)$ for band A and B are shown in Fig. \ref{transitions2}, in comparison with the data \cite{Cs130-ZhuLH}. Fig. \ref{transitions2}(a) shows that the intra band $B(E2)$ are similar for bands A and B, in both experimental and theoretical results, as expected for a pair of chiral doublets \cite{Koike2004,WSY2007}. The $B(E2)$ values are somewhat overestimated by the calculation. The similarity of the intraband $B(M1)$ between the chiral doublets, together with the staggering behavior of both intra- and interband $B(M1)$, are found in Figs. \ref{transitions2}(b) and (c), in both experimental and theoretical results. These features are also expected as signatures of the chiral modes according to Refs. \cite{Koike2004,WSY2007}. The characteristic features of the electromagnetic transitions expected for the chiral doublet bands shown in Fig. \ref{transitions2}, which have been reproduced with a single configuration \cite{CFQ2017}, are persistent even when the configuration mixing is taken into account.

%\begin{figure}[!h]
%  \begin{center}
%    \includegraphics[width=10 cm]{S(I)-ABCD.eps}
%    \caption{(Color online) The values of $S(I)=[E(I)-E(I-1)]/2I$ for the bands A,B and bands C, D, compared with the experimental values.}\label{S(I)-ABCD}
%  \end{center}
%\end{figure}

\begin{figure}[!h]
  \begin{center}
    \includegraphics[width=15 cm]{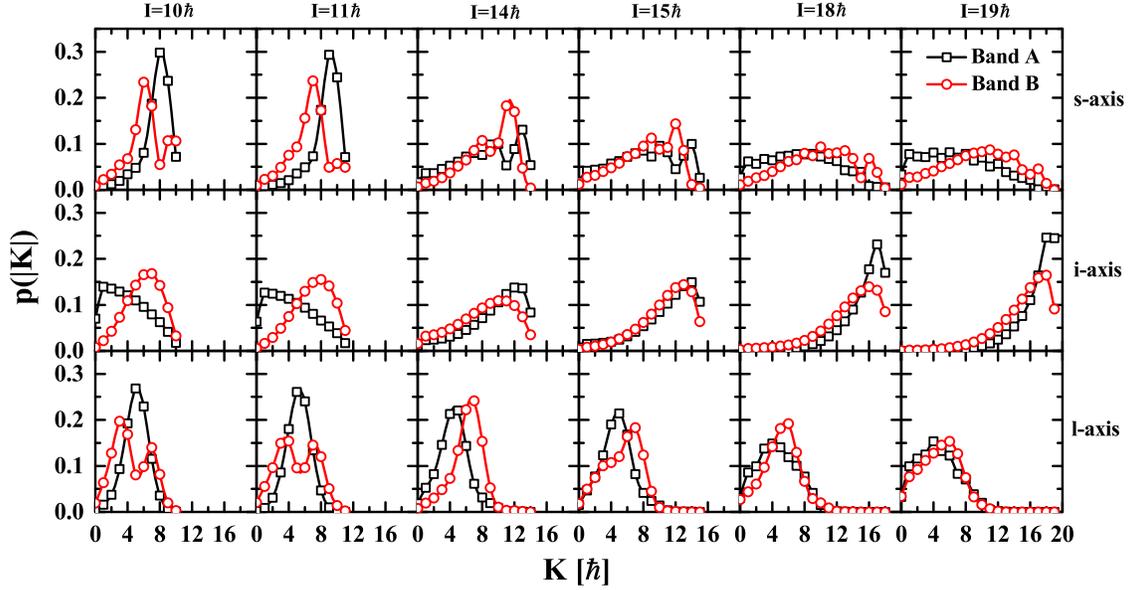}
    \caption{(Color online) The $K~plot$, i.e., the $K$ distributions for the angular momentum on the short ($s$), intermediate ($i$) and long ($l$) axes, calculated at spins $I=10, 11, 14, 15, 18, 19~\hbar$, respectively, for the chiral doublet bands in $^{130}$Cs. }\label{Kdistributions-1208}
  \end{center}
\end{figure}

\begin{figure}[!h]
  \begin{center}
    \includegraphics[width=17 cm]{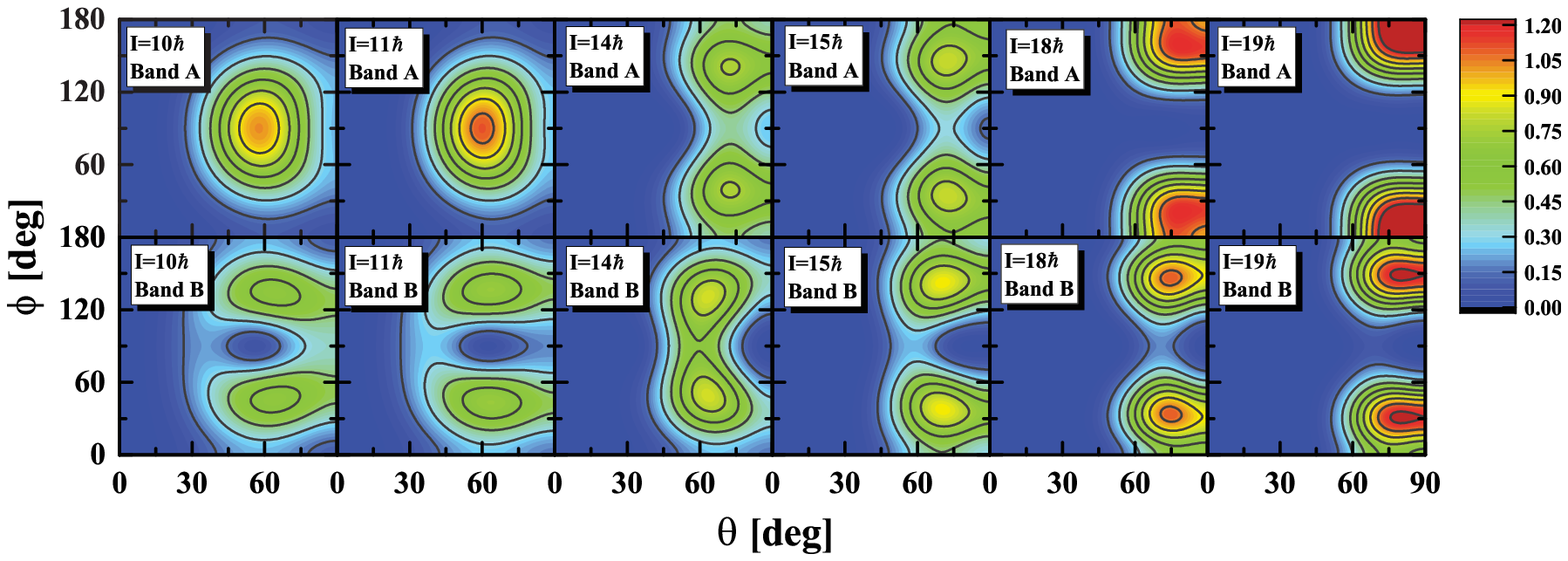}
    \caption{(Color online) The $azimuthal~plot$, i.e., profile for the orientation of the angular momentum on the $(\theta,\phi)$ plane, calculated at spins $I=10, 11, 14, 15, 18, 19~\hbar$, respectively, for the chiral doublet bands in $^{130}$Cs.}\label{tiltedangles-AB}
  \end{center}
\end{figure}

The angular momentum geometry of the chiral doublets A and B can be illustrated by the $K~plot$ and the $azimuthal~plot$, as in Ref. \cite{CFQ2017}. The $K~plot$s, i.e., the $K$ distributions $p^{I\sigma}(|K|)$ for the angular momentum on the three principle axes, are shown in Fig. \ref{Kdistributions-1208} at the spins $I=10, 11, 14, 15, 18, 19~\hbar$. These results could be compared with the corresponding results obtained by the PRM, which have been discussed in Refs. \cite{QB2009PLB,QB2009PRC}.

As seen in the $K~plot$ in Fig. \ref{Kdistributions-1208}, the evolution of the chiral modes can be exhibited.

For spins $I=10$ and $11~\hbar$, the $K~plots$ shown in Fig. \ref{Kdistributions-1208} are in accordance with the expectation for chiral vibration with respect to the $s$-$l$ plane, which takes place near the band head with insufficient collective rotation. The probability at $K_i=0$ is significant for band A, which indicates a wave function symmetric with respect to $K_i=0$ and corresponds to a 0-phonon state. On the other hand, the vanishing probability at $K_i=0$ for band B indicates an antisymmetric wave function corresponding to a 1-phonon state.

For spins $I=14$ and $15~\hbar$, the most probable value for $K_i$ appears at $K_i\sim13~\hbar$ for band A, which suggests that the collective rotation around the $i$-axis develops and the angular momenta for band A deviate from the $s$-$l$ plane with the increase of spin. The peaks of the $K_l$- and $K_s$-distributions also correspond to $K$ values substantially away from zero, indicating the occurrence of static chirality. The $K~plot$s for band B are similar to those for band A, and the occurrence of static chirality is thus supported. The tunneling between the left-handed and the right-handed configurations is responsible for the energy separation between the two bands, which explains why the doublets are closest to each other in energy at $I=14$ and $15~\hbar$.

For spins $I=18$ and $19~\hbar$, the $K_s$- and $K_l$-distributions become broad, and the peaks of the $K_i$-distributions become sharp at $K_i\sim I$. Both features suggest that the angular momenta move close to the $i$-axis. Therefore the static chirality disappears and the aplanar rotation is replaced by the principle axis rotation.

In Fig. \ref{tiltedangles-AB}, the profiles for the
orientation of the angular momentum on the $(\theta,\phi)$ plane, the $azimuthal~plot$s, are shown for the same spins as in Fig. \ref{Kdistributions-1208}. The definitions of the angles $(\theta,\phi)$ can be found in Ref. \cite{CFQ2017}.

For spins $I=10$ and $11~\hbar$, the profiles for the
orientation of the angular momentum for band A have a single peak at $(\theta\sim 60^\circ, \phi=90^\circ)$, which suggests that the angular momentum stays within the $s$-$l$ plane, in accordance with the expectation for a 0-phonon state. On the other hand, the profiles for band B show a node at $(\theta\sim 60^\circ, \phi=90^\circ)$, with two peaks at $(\theta\sim65^\circ,\phi\sim45^\circ)$ and $(\theta\sim65^\circ,\phi\sim135^\circ)$, respectively. The existence of the node and the two peaks supports the interpretation of a 1-phonon vibration. Therefore the interpretation of chiral vibration is demonstrated.

For spins $I=14$ and $15~\hbar$, the $azimuthal~plot$s for bands A and B are similar. Two peaks corresponding to aplanar orientations are found, i.e. $(\theta\sim75^\circ,\phi\sim35^\circ)$ and $(\theta\sim75^\circ,\phi\sim145^\circ)$ for band A, while $(\theta\sim70^\circ,\phi\sim40^\circ)$ and $(\theta\sim70^\circ,\phi\sim140^\circ)$ for band B. These features could be understood as a realization of static chirality. The non-vanishing distribution for $\theta=90^\circ$ and $\phi=90^\circ$ reflects the tunneling between the left- and right-handed configurations.

For spins $I=18$ and $19~\hbar$, the peaks for the $azimuthal~plot$s for band A move toward ($\theta\sim80^\circ,\phi\sim20^\circ$) and ($\theta\sim80^\circ,\phi\sim160^\circ$), namely close to the $i$-axis. This is in accordance with the dominance of rotation around the $i$-axis reflected in the $K~plot$ discussed above, suggesting the disappearance of chiral geometry and the onset of principle axis rotation. The peaks for the $azimuthal~plot$s for band B locate at $(\theta\sim75^\circ,\phi\sim35^\circ)$ and $(\theta\sim75^\circ,\phi\sim145^\circ)$, which are similar to those at $I=14,15~\hbar$ but approaching the $i$-axis. Thus the chiral geometry is weakened. In general, the evolution of the angular momentum geometry in the $azimuthal~plot$ is consistent with those in the $K~plot$.

By blocking
the lowest $\pi h_{11/2}$ and the fourth $\nu h_{11/2}$ orbitals, the chiral geometry in $K~plot$ and $azimuthal~plot$ has been discussed in Ref. \cite{CFQ2017}. Figures \ref{Kdistributions-1208} and \ref{tiltedangles-AB} confirm that the chiral geometry in $K~plot$ and $azimuthal~plot$ can be stable against configuration mixing.

%\subsection{SUMMARY}

%In this work the projected shell model with configuration mixing for nuclear chirality is developed and applied to $^{130}$Cs \cite{Starosta2001}. The chiral geometry for the chiral doublet bands is illustrated in terms of the $K~plot$ and the $azimuthal~plot$, in a similar way as Ref. \cite{CFQ2017}. The effect of spin alignments on the angular momentum geometry and the description of the other rotational bands are discussed.

In summary, the projected shell model with configuration mixing for nuclear chirality is developed and applied to the observed rotational bands in the chiral nucleus $^{130}$Cs. Both the chiral and achiral bands are described on the same foot.
For the chiral bands, the energy spectra, $S(I)$, $B(M1)$ and $B(E2)$ are well reproduced. The chiral geometry is demonstrated in the $K~plot$ and the $azithumal~plot$. The stability of the chiral geometry against the configuration mixing is confirmed. As another advantage, the other rotational bands are described simultaneously with the same Hamiltonian.

%\section*{Acknowledgements}

%We thank Z. C. Gao for his help and discussion in the implementation of the computing codes, and X. H. Wu for his help in preparing Figure \ref{figangles}. Stimulating discussions with Z. C. Gao, F. Pan, Y. Sun, S. Y. Wang and other paticipants in the workshop ``Nuclear spontaneous symmetry breaking and its experimental signals" are acknowledged.

This work was partly supported by the Chinese Major State 973 Program No. 2018YFA0404400, the National Natural Science Foundation of China (Grants No. 11335002 and No. 11621131001), and the China Postdoctoral Science Foundation under Grant No. 2017M610688.

\bibliography{mybib}

\begin{thebibliography}{58}
\expandafter\ifx\csname natexlab\endcsname\relax\def\natexlab#1{#1}\fi
\expandafter\ifx\csname bibnamefont\endcsname\relax
  \def\bibnamefont#1{#1}\fi
\expandafter\ifx\csname bibfnamefont\endcsname\relax
  \def\bibfnamefont#1{#1}\fi
\expandafter\ifx\csname citenamefont\endcsname\relax
  \def\citenamefont#1{#1}\fi
\expandafter\ifx\csname url\endcsname\relax
  \def\url#1{\texttt{#1}}\fi
\expandafter\ifx\csname urlprefix\endcsname\relax\def\urlprefix{URL }\fi
\providecommand{\bibinfo}[2]{#2}
\providecommand{\eprint}[2][]{\url{#2}}

\bibitem[{\citenamefont{Frauendorf and Meng}(1997)}]{TAC1997}
\bibinfo{author}{\bibfnamefont{S.}~\bibnamefont{Frauendorf}} \bibnamefont{and}
  \bibinfo{author}{\bibfnamefont{J.}~\bibnamefont{Meng}},
  \bibinfo{journal}{Nucl. Phys. A} \textbf{\bibinfo{volume}{617}},
  \bibinfo{pages}{131 } (\bibinfo{year}{1997}).

\bibitem[{\citenamefont{Wang et~al.}(2011)\citenamefont{Wang, Qi, Liu, Zhang,
  Hua, Li, Chen, Zhu, Meng, Wyngaardt et~al.}}]{Br80exp}
\bibinfo{author}{\bibfnamefont{S.~Y.} \bibnamefont{Wang}},
  \bibinfo{author}{\bibfnamefont{B.}~\bibnamefont{Qi}},
  \bibinfo{author}{\bibfnamefont{L.}~\bibnamefont{Liu}},
  \bibinfo{author}{\bibfnamefont{S.~Q.} \bibnamefont{Zhang}},
  \bibinfo{author}{\bibfnamefont{H.}~\bibnamefont{Hua}},
  \bibinfo{author}{\bibfnamefont{X.~Q.} \bibnamefont{Li}},
  \bibinfo{author}{\bibfnamefont{Y.~Y.} \bibnamefont{Chen}},
  \bibinfo{author}{\bibfnamefont{L.~H.} \bibnamefont{Zhu}},
  \bibinfo{author}{\bibfnamefont{J.}~\bibnamefont{Meng}},
  \bibinfo{author}{\bibfnamefont{S.~M.} \bibnamefont{Wyngaardt}},
  \bibnamefont{et~al.}, \bibinfo{journal}{Phys. Lett. B}
  \textbf{\bibinfo{volume}{703}}, \bibinfo{pages}{40 } (\bibinfo{year}{2011}).

\bibitem[{\citenamefont{Liu et~al.}(2016)\citenamefont{Liu, Wang, Bark, Zhang,
  Meng, Qi, Jones, Wyngaardt, Zhao, Xu et~al.}}]{Br78exp}
\bibinfo{author}{\bibfnamefont{C.}~\bibnamefont{Liu}},
  \bibinfo{author}{\bibfnamefont{S.~Y.} \bibnamefont{Wang}},
  \bibinfo{author}{\bibfnamefont{R.~A.} \bibnamefont{Bark}},
  \bibinfo{author}{\bibfnamefont{S.~Q.} \bibnamefont{Zhang}},
  \bibinfo{author}{\bibfnamefont{J.}~\bibnamefont{Meng}},
  \bibinfo{author}{\bibfnamefont{B.}~\bibnamefont{Qi}},
  \bibinfo{author}{\bibfnamefont{P.}~\bibnamefont{Jones}},
  \bibinfo{author}{\bibfnamefont{S.~M.} \bibnamefont{Wyngaardt}},
  \bibinfo{author}{\bibfnamefont{J.}~\bibnamefont{Zhao}},
  \bibinfo{author}{\bibfnamefont{C.}~\bibnamefont{Xu}}, \bibnamefont{et~al.},
  \bibinfo{journal}{Phys. Rev. Lett.} \textbf{\bibinfo{volume}{116}},
  \bibinfo{pages}{112501} (\bibinfo{year}{2016}).

\bibitem[{\citenamefont{Vaman et~al.}(2004)\citenamefont{Vaman, Fossan, Koike,
  Starosta, Lee, and Macchiavelli}}]{Rh104exp}
\bibinfo{author}{\bibfnamefont{C.}~\bibnamefont{Vaman}},
  \bibinfo{author}{\bibfnamefont{D.~B.} \bibnamefont{Fossan}},
  \bibinfo{author}{\bibfnamefont{T.}~\bibnamefont{Koike}},
  \bibinfo{author}{\bibfnamefont{K.}~\bibnamefont{Starosta}},
  \bibinfo{author}{\bibfnamefont{I.~Y.} \bibnamefont{Lee}}, \bibnamefont{and}
  \bibinfo{author}{\bibfnamefont{A.~O.} \bibnamefont{Macchiavelli}},
  \bibinfo{journal}{Phys. Rev. Lett.} \textbf{\bibinfo{volume}{92}},
  \bibinfo{pages}{032501} (\bibinfo{year}{2004}).

\bibitem[{\citenamefont{Joshi et~al.}(2007)\citenamefont{Joshi, Carpenter,
  Fossan, Koike, Paul, Rainovski, Starosta, Vaman, and Wadsworth}}]{Ag106exp1}
\bibinfo{author}{\bibfnamefont{P.}~\bibnamefont{Joshi}},
  \bibinfo{author}{\bibfnamefont{M.~P.} \bibnamefont{Carpenter}},
  \bibinfo{author}{\bibfnamefont{D.~B.} \bibnamefont{Fossan}},
  \bibinfo{author}{\bibfnamefont{T.}~\bibnamefont{Koike}},
  \bibinfo{author}{\bibfnamefont{E.~S.} \bibnamefont{Paul}},
  \bibinfo{author}{\bibfnamefont{G.}~\bibnamefont{Rainovski}},
  \bibinfo{author}{\bibfnamefont{K.}~\bibnamefont{Starosta}},
  \bibinfo{author}{\bibfnamefont{C.}~\bibnamefont{Vaman}}, \bibnamefont{and}
  \bibinfo{author}{\bibfnamefont{R.}~\bibnamefont{Wadsworth}},
  \bibinfo{journal}{Phys. Rev. Lett.} \textbf{\bibinfo{volume}{98}},
  \bibinfo{pages}{102501} (\bibinfo{year}{2007}).

\bibitem[{\citenamefont{Tonev et~al.}(2014)\citenamefont{Tonev, Yavahchova,
  Goutev, de~Angelis, Petkov, Bhowmik, Singh, Muralithar, Madhavan, Kumar
  et~al.}}]{Rh102exp}
\bibinfo{author}{\bibfnamefont{D.}~\bibnamefont{Tonev}},
  \bibinfo{author}{\bibfnamefont{M.~S.} \bibnamefont{Yavahchova}},
  \bibinfo{author}{\bibfnamefont{N.}~\bibnamefont{Goutev}},
  \bibinfo{author}{\bibfnamefont{G.}~\bibnamefont{de~Angelis}},
  \bibinfo{author}{\bibfnamefont{P.}~\bibnamefont{Petkov}},
  \bibinfo{author}{\bibfnamefont{R.~K.} \bibnamefont{Bhowmik}},
  \bibinfo{author}{\bibfnamefont{R.~P.} \bibnamefont{Singh}},
  \bibinfo{author}{\bibfnamefont{S.}~\bibnamefont{Muralithar}},
  \bibinfo{author}{\bibfnamefont{N.}~\bibnamefont{Madhavan}},
  \bibinfo{author}{\bibfnamefont{R.}~\bibnamefont{Kumar}},
  \bibnamefont{et~al.}, \bibinfo{journal}{Phys. Rev. Lett.}
  \textbf{\bibinfo{volume}{112}}, \bibinfo{pages}{052501}
  (\bibinfo{year}{2014}).

\bibitem[{\citenamefont{Lieder et~al.}(2014)\citenamefont{Lieder, Lieder, Bark,
  Chen, Zhang, Meng, Lawrie, Lawrie, Bvumbi, Kheswa et~al.}}]{Ag106exp3}
\bibinfo{author}{\bibfnamefont{E.~O.} \bibnamefont{Lieder}},
  \bibinfo{author}{\bibfnamefont{R.~M.} \bibnamefont{Lieder}},
  \bibinfo{author}{\bibfnamefont{R.~A.} \bibnamefont{Bark}},
  \bibinfo{author}{\bibfnamefont{Q.~B.} \bibnamefont{Chen}},
  \bibinfo{author}{\bibfnamefont{S.~Q.} \bibnamefont{Zhang}},
  \bibinfo{author}{\bibfnamefont{J.}~\bibnamefont{Meng}},
  \bibinfo{author}{\bibfnamefont{E.~A.} \bibnamefont{Lawrie}},
  \bibinfo{author}{\bibfnamefont{J.~J.} \bibnamefont{Lawrie}},
  \bibinfo{author}{\bibfnamefont{S.~P.} \bibnamefont{Bvumbi}},
  \bibinfo{author}{\bibfnamefont{N.~Y.} \bibnamefont{Kheswa}},
  \bibnamefont{et~al.}, \bibinfo{journal}{Phys. Rev. Lett.}
  \textbf{\bibinfo{volume}{112}}, \bibinfo{pages}{202502}
  (\bibinfo{year}{2014}).

\bibitem[{\citenamefont{Rather et~al.}(2014)\citenamefont{Rather, Datta,
  Chattopadhyay, Rajbanshi, Goswami, Bhat, Sheikh, Roy, Palit, Pal
  et~al.}}]{Ag106exp2}
\bibinfo{author}{\bibfnamefont{N.}~\bibnamefont{Rather}},
  \bibinfo{author}{\bibfnamefont{P.}~\bibnamefont{Datta}},
  \bibinfo{author}{\bibfnamefont{S.}~\bibnamefont{Chattopadhyay}},
  \bibinfo{author}{\bibfnamefont{S.}~\bibnamefont{Rajbanshi}},
  \bibinfo{author}{\bibfnamefont{A.}~\bibnamefont{Goswami}},
  \bibinfo{author}{\bibfnamefont{G.~H.} \bibnamefont{Bhat}},
  \bibinfo{author}{\bibfnamefont{J.~A.} \bibnamefont{Sheikh}},
  \bibinfo{author}{\bibfnamefont{S.}~\bibnamefont{Roy}},
  \bibinfo{author}{\bibfnamefont{R.}~\bibnamefont{Palit}},
  \bibinfo{author}{\bibfnamefont{S.}~\bibnamefont{Pal}}, \bibnamefont{et~al.},
  \bibinfo{journal}{Phys. Rev. Lett.} \textbf{\bibinfo{volume}{112}},
  \bibinfo{pages}{202503} (\bibinfo{year}{2014}).

\bibitem[{\citenamefont{Kuti et~al.}(2014)\citenamefont{Kuti, Chen, Tim\'ar,
  Sohler, Zhang, Zhang, Zhao, Meng, Starosta, Koike et~al.}}]{Rh103exp}
\bibinfo{author}{\bibfnamefont{I.}~\bibnamefont{Kuti}},
  \bibinfo{author}{\bibfnamefont{Q.~B.} \bibnamefont{Chen}},
  \bibinfo{author}{\bibfnamefont{J.}~\bibnamefont{Tim\'ar}},
  \bibinfo{author}{\bibfnamefont{D.}~\bibnamefont{Sohler}},
  \bibinfo{author}{\bibfnamefont{S.~Q.} \bibnamefont{Zhang}},
  \bibinfo{author}{\bibfnamefont{Z.~H.} \bibnamefont{Zhang}},
  \bibinfo{author}{\bibfnamefont{P.~W.} \bibnamefont{Zhao}},
  \bibinfo{author}{\bibfnamefont{J.}~\bibnamefont{Meng}},
  \bibinfo{author}{\bibfnamefont{K.}~\bibnamefont{Starosta}},
  \bibinfo{author}{\bibfnamefont{T.}~\bibnamefont{Koike}},
  \bibnamefont{et~al.}, \bibinfo{journal}{Phys. Rev. Lett.}
  \textbf{\bibinfo{volume}{113}}, \bibinfo{pages}{032501}
  (\bibinfo{year}{2014}).

\bibitem[{\citenamefont{Starosta et~al.}(2001)\citenamefont{Starosta, Koike,
  Chiara, Fossan, LaFosse, Hecht, Beausang, Caprio, Cooper, Kr\"ucken
  et~al.}}]{Starosta2001}
\bibinfo{author}{\bibfnamefont{K.}~\bibnamefont{Starosta}},
  \bibinfo{author}{\bibfnamefont{T.}~\bibnamefont{Koike}},
  \bibinfo{author}{\bibfnamefont{C.~J.} \bibnamefont{Chiara}},
  \bibinfo{author}{\bibfnamefont{D.~B.} \bibnamefont{Fossan}},
  \bibinfo{author}{\bibfnamefont{D.~R.} \bibnamefont{LaFosse}},
  \bibinfo{author}{\bibfnamefont{A.~A.} \bibnamefont{Hecht}},
  \bibinfo{author}{\bibfnamefont{C.~W.} \bibnamefont{Beausang}},
  \bibinfo{author}{\bibfnamefont{M.~A.} \bibnamefont{Caprio}},
  \bibinfo{author}{\bibfnamefont{J.~R.} \bibnamefont{Cooper}},
  \bibinfo{author}{\bibfnamefont{R.}~\bibnamefont{Kr\"ucken}},
  \bibnamefont{et~al.}, \bibinfo{journal}{Phys. Rev. Lett.}
  \textbf{\bibinfo{volume}{86}}, \bibinfo{pages}{971} (\bibinfo{year}{2001}).

\bibitem[{\citenamefont{Zhu et~al.}(2003)\citenamefont{Zhu, Garg, Nayak,
  Ghugre, Pattabiraman, Fossan, Koike, Starosta, Vaman, Janssens
  et~al.}}]{Nd135exp}
\bibinfo{author}{\bibfnamefont{S.}~\bibnamefont{Zhu}},
  \bibinfo{author}{\bibfnamefont{U.}~\bibnamefont{Garg}},
  \bibinfo{author}{\bibfnamefont{B.~K.} \bibnamefont{Nayak}},
  \bibinfo{author}{\bibfnamefont{S.~S.} \bibnamefont{Ghugre}},
  \bibinfo{author}{\bibfnamefont{N.~S.} \bibnamefont{Pattabiraman}},
  \bibinfo{author}{\bibfnamefont{D.~B.} \bibnamefont{Fossan}},
  \bibinfo{author}{\bibfnamefont{T.}~\bibnamefont{Koike}},
  \bibinfo{author}{\bibfnamefont{K.}~\bibnamefont{Starosta}},
  \bibinfo{author}{\bibfnamefont{C.}~\bibnamefont{Vaman}},
  \bibinfo{author}{\bibfnamefont{R.~V.~F.} \bibnamefont{Janssens}},
  \bibnamefont{et~al.}, \bibinfo{journal}{Phys. Rev. Lett.}
  \textbf{\bibinfo{volume}{91}}, \bibinfo{pages}{132501}
  (\bibinfo{year}{2003}).

\bibitem[{\citenamefont{Tonev et~al.}(2006)\citenamefont{Tonev, de~Angelis,
  Petkov, Dewald, Brant, Frauendorf, Balabanski, Pejovic, Bazzacco, Bednarczyk
  et~al.}}]{Pr134exp1}
\bibinfo{author}{\bibfnamefont{D.}~\bibnamefont{Tonev}},
  \bibinfo{author}{\bibfnamefont{G.}~\bibnamefont{de~Angelis}},
  \bibinfo{author}{\bibfnamefont{P.}~\bibnamefont{Petkov}},
  \bibinfo{author}{\bibfnamefont{A.}~\bibnamefont{Dewald}},
  \bibinfo{author}{\bibfnamefont{S.}~\bibnamefont{Brant}},
  \bibinfo{author}{\bibfnamefont{S.}~\bibnamefont{Frauendorf}},
  \bibinfo{author}{\bibfnamefont{D.~L.} \bibnamefont{Balabanski}},
  \bibinfo{author}{\bibfnamefont{P.}~\bibnamefont{Pejovic}},
  \bibinfo{author}{\bibfnamefont{D.}~\bibnamefont{Bazzacco}},
  \bibinfo{author}{\bibfnamefont{P.}~\bibnamefont{Bednarczyk}},
  \bibnamefont{et~al.}, \bibinfo{journal}{Phys. Rev. Lett.}
  \textbf{\bibinfo{volume}{96}}, \bibinfo{pages}{052501}
  (\bibinfo{year}{2006}).

\bibitem[{\citenamefont{Petrache et~al.}(2006)\citenamefont{Petrache, Hagemann,
  Hamamoto, and Starosta}}]{Pr134exp2}
\bibinfo{author}{\bibfnamefont{C.~M.} \bibnamefont{Petrache}},
  \bibinfo{author}{\bibfnamefont{G.~B.} \bibnamefont{Hagemann}},
  \bibinfo{author}{\bibfnamefont{I.}~\bibnamefont{Hamamoto}}, \bibnamefont{and}
  \bibinfo{author}{\bibfnamefont{K.}~\bibnamefont{Starosta}},
  \bibinfo{journal}{Phys. Rev. Lett.} \textbf{\bibinfo{volume}{96}},
  \bibinfo{pages}{112502} (\bibinfo{year}{2006}).

\bibitem[{\citenamefont{Grodner et~al.}(2006)\citenamefont{Grodner, Srebrny,
  Pasternak, Zalewska, Morek, Droste, Mierzejewski, Kowalczyk, Kownacki,
  Kisieli{\'{n}}ski et~al.}}]{Cs128-2006}
\bibinfo{author}{\bibfnamefont{E.}~\bibnamefont{Grodner}},
  \bibinfo{author}{\bibfnamefont{J.}~\bibnamefont{Srebrny}},
  \bibinfo{author}{\bibfnamefont{A.~A.} \bibnamefont{Pasternak}},
  \bibinfo{author}{\bibfnamefont{I.}~\bibnamefont{Zalewska}},
  \bibinfo{author}{\bibfnamefont{T.}~\bibnamefont{Morek}},
  \bibinfo{author}{\bibfnamefont{C.}~\bibnamefont{Droste}},
  \bibinfo{author}{\bibfnamefont{J.}~\bibnamefont{Mierzejewski}},
  \bibinfo{author}{\bibfnamefont{M.}~\bibnamefont{Kowalczyk}},
  \bibinfo{author}{\bibfnamefont{J.}~\bibnamefont{Kownacki}},
  \bibinfo{author}{\bibfnamefont{M.}~\bibnamefont{Kisieli{\'{n}}ski}},
  \bibnamefont{et~al.}, \bibinfo{journal}{Phys. Rev. Lett.}
  \textbf{\bibinfo{volume}{97}}, \bibinfo{pages}{172501}
  (\bibinfo{year}{2006}).

\bibitem[{\citenamefont{Mukhopadhyay et~al.}(2007)\citenamefont{Mukhopadhyay,
  Almehed, Garg, Frauendorf, Li, Rao, Wang, Ghugre, Carpenter, Gros
  et~al.}}]{Nd135exp2}
\bibinfo{author}{\bibfnamefont{S.}~\bibnamefont{Mukhopadhyay}},
  \bibinfo{author}{\bibfnamefont{D.}~\bibnamefont{Almehed}},
  \bibinfo{author}{\bibfnamefont{U.}~\bibnamefont{Garg}},
  \bibinfo{author}{\bibfnamefont{S.}~\bibnamefont{Frauendorf}},
  \bibinfo{author}{\bibfnamefont{T.}~\bibnamefont{Li}},
  \bibinfo{author}{\bibfnamefont{P.~V.~M.} \bibnamefont{Rao}},
  \bibinfo{author}{\bibfnamefont{X.}~\bibnamefont{Wang}},
  \bibinfo{author}{\bibfnamefont{S.~S.} \bibnamefont{Ghugre}},
  \bibinfo{author}{\bibfnamefont{M.~P.} \bibnamefont{Carpenter}},
  \bibinfo{author}{\bibfnamefont{S.}~\bibnamefont{Gros}}, \bibnamefont{et~al.},
  \bibinfo{journal}{Phys. Rev. Lett.} \textbf{\bibinfo{volume}{99}},
  \bibinfo{pages}{172501} (\bibinfo{year}{2007}).

\bibitem[{\citenamefont{Ayangeakaa et~al.}(2013)\citenamefont{Ayangeakaa, Garg,
  Anthony, Frauendorf, Matta, Nayak, Patel, Chen, Zhang, Zhao
  et~al.}}]{Ce133exp}
\bibinfo{author}{\bibfnamefont{A.~D.} \bibnamefont{Ayangeakaa}},
  \bibinfo{author}{\bibfnamefont{U.}~\bibnamefont{Garg}},
  \bibinfo{author}{\bibfnamefont{M.~D.} \bibnamefont{Anthony}},
  \bibinfo{author}{\bibfnamefont{S.}~\bibnamefont{Frauendorf}},
  \bibinfo{author}{\bibfnamefont{J.~T.} \bibnamefont{Matta}},
  \bibinfo{author}{\bibfnamefont{B.~K.} \bibnamefont{Nayak}},
  \bibinfo{author}{\bibfnamefont{D.}~\bibnamefont{Patel}},
  \bibinfo{author}{\bibfnamefont{Q.~B.} \bibnamefont{Chen}},
  \bibinfo{author}{\bibfnamefont{S.~Q.} \bibnamefont{Zhang}},
  \bibinfo{author}{\bibfnamefont{P.~W.} \bibnamefont{Zhao}},
  \bibnamefont{et~al.}, \bibinfo{journal}{Phys. Rev. Lett.}
  \textbf{\bibinfo{volume}{110}}, \bibinfo{pages}{172504}
  (\bibinfo{year}{2013}).

\bibitem[{\citenamefont{Balabanski et~al.}(2004)\citenamefont{Balabanski,
  Danchev, Hartley, Riedinger, Zeidan, Zhang, Barton, Beausang, Caprio, Casten
  et~al.}}]{Ir188exp}
\bibinfo{author}{\bibfnamefont{D.~L.} \bibnamefont{Balabanski}},
  \bibinfo{author}{\bibfnamefont{M.}~\bibnamefont{Danchev}},
  \bibinfo{author}{\bibfnamefont{D.~J.} \bibnamefont{Hartley}},
  \bibinfo{author}{\bibfnamefont{L.~L.} \bibnamefont{Riedinger}},
  \bibinfo{author}{\bibfnamefont{O.}~\bibnamefont{Zeidan}},
  \bibinfo{author}{\bibfnamefont{J.~Y.} \bibnamefont{Zhang}},
  \bibinfo{author}{\bibfnamefont{C.~J.} \bibnamefont{Barton}},
  \bibinfo{author}{\bibfnamefont{C.~W.} \bibnamefont{Beausang}},
  \bibinfo{author}{\bibfnamefont{M.~A.} \bibnamefont{Caprio}},
  \bibinfo{author}{\bibfnamefont{R.~F.} \bibnamefont{Casten}},
  \bibnamefont{et~al.}, \bibinfo{journal}{Phys. Rev. C}
  \textbf{\bibinfo{volume}{70}}, \bibinfo{pages}{044305}
  (\bibinfo{year}{2004}).

\bibitem[{\citenamefont{Lawrie et~al.}(2008)\citenamefont{Lawrie, Vymers,
  Lawrie, Vieu, Bark, Lindsay, Mabala, Maliage, Masiteng, Mullins
  et~al.}}]{Tl198exp}
\bibinfo{author}{\bibfnamefont{E.~A.} \bibnamefont{Lawrie}},
  \bibinfo{author}{\bibfnamefont{P.~A.} \bibnamefont{Vymers}},
  \bibinfo{author}{\bibfnamefont{J.~J.} \bibnamefont{Lawrie}},
  \bibinfo{author}{\bibfnamefont{C.}~\bibnamefont{Vieu}},
  \bibinfo{author}{\bibfnamefont{R.~A.} \bibnamefont{Bark}},
  \bibinfo{author}{\bibfnamefont{R.}~\bibnamefont{Lindsay}},
  \bibinfo{author}{\bibfnamefont{G.~K.} \bibnamefont{Mabala}},
  \bibinfo{author}{\bibfnamefont{S.~M.} \bibnamefont{Maliage}},
  \bibinfo{author}{\bibfnamefont{P.~L.} \bibnamefont{Masiteng}},
  \bibinfo{author}{\bibfnamefont{S.~M.} \bibnamefont{Mullins}},
  \bibnamefont{et~al.}, \bibinfo{journal}{Phys. Rev. C}
  \textbf{\bibinfo{volume}{78}}, \bibinfo{pages}{021305(R)}
  (\bibinfo{year}{2008}).

\bibitem[{\citenamefont{Meng and Zhang}(2010)}]{Meng2010JPG}
\bibinfo{author}{\bibfnamefont{J.}~\bibnamefont{Meng}} \bibnamefont{and}
  \bibinfo{author}{\bibfnamefont{S.~Q.} \bibnamefont{Zhang}},
  \bibinfo{journal}{J. Phys. G: Nucl. Part. Phys.}
  \textbf{\bibinfo{volume}{37}}, \bibinfo{pages}{064025}
  (\bibinfo{year}{2010}).

\bibitem[{\citenamefont{Meng et~al.}(2014)\citenamefont{Meng, Chen, and
  Zhang}}]{Meng2014IJMPE}
\bibinfo{author}{\bibfnamefont{J.}~\bibnamefont{Meng}},
  \bibinfo{author}{\bibfnamefont{Q.~B.} \bibnamefont{Chen}}, \bibnamefont{and}
  \bibinfo{author}{\bibfnamefont{S.~Q.} \bibnamefont{Zhang}},
  \bibinfo{journal}{Int. J. Mod. Phys. E} \textbf{\bibinfo{volume}{23}},
  \bibinfo{pages}{1430016} (\bibinfo{year}{2014}).

\bibitem[{\citenamefont{Meng and Zhao}(2016)}]{Meng2016PS}
\bibinfo{author}{\bibfnamefont{J.}~\bibnamefont{Meng}} \bibnamefont{and}
  \bibinfo{author}{\bibfnamefont{P.~W.} \bibnamefont{Zhao}},
  \bibinfo{journal}{Phys. Scr.} \textbf{\bibinfo{volume}{91}},
  \bibinfo{pages}{053008} (\bibinfo{year}{2016}).

\bibitem[{\citenamefont{Xiong and Wang}(2018)}]{Xiong2018}
\bibinfo{author}{\bibfnamefont{B.~W.} \bibnamefont{Xiong}} \bibnamefont{and}
  \bibinfo{author}{\bibfnamefont{Y.~Y.} \bibnamefont{Wang}},
  \bibinfo{journal}{arXiv:1804.04437}  (\bibinfo{year}{2018}).

\bibitem[{\citenamefont{Peng et~al.}(2003)\citenamefont{Peng, Meng, and
  Zhang}}]{Pengjing2003}
\bibinfo{author}{\bibfnamefont{J.}~\bibnamefont{Peng}},
  \bibinfo{author}{\bibfnamefont{J.}~\bibnamefont{Meng}}, \bibnamefont{and}
  \bibinfo{author}{\bibfnamefont{S.~Q.} \bibnamefont{Zhang}},
  \bibinfo{journal}{Phys. Rev. C} \textbf{\bibinfo{volume}{68}},
  \bibinfo{pages}{044324} (\bibinfo{year}{2003}).

\bibitem[{\citenamefont{Koike et~al.}(2004)\citenamefont{Koike, Starosta, and
  Hamamoto}}]{Koike2004}
\bibinfo{author}{\bibfnamefont{T.}~\bibnamefont{Koike}},
  \bibinfo{author}{\bibfnamefont{K.}~\bibnamefont{Starosta}}, \bibnamefont{and}
  \bibinfo{author}{\bibfnamefont{I.}~\bibnamefont{Hamamoto}},
  \bibinfo{journal}{Phys. Rev. Lett.} \textbf{\bibinfo{volume}{93}},
  \bibinfo{pages}{172502} (\bibinfo{year}{2004}).

\bibitem[{\citenamefont{Zhang et~al.}(2007)\citenamefont{Zhang, Qi, Wang, and
  Meng}}]{ZSQ2007}
\bibinfo{author}{\bibfnamefont{S.~Q.} \bibnamefont{Zhang}},
  \bibinfo{author}{\bibfnamefont{B.}~\bibnamefont{Qi}},
  \bibinfo{author}{\bibfnamefont{S.~Y.} \bibnamefont{Wang}}, \bibnamefont{and}
  \bibinfo{author}{\bibfnamefont{J.}~\bibnamefont{Meng}},
  \bibinfo{journal}{Phys. Rev. C} \textbf{\bibinfo{volume}{75}},
  \bibinfo{pages}{044307} (\bibinfo{year}{2007}).

\bibitem[{\citenamefont{Qi et~al.}(2009{\natexlab{a}})\citenamefont{Qi, Zhang,
  Meng, Wang, and Frauendorf}}]{QB2009PLB}
\bibinfo{author}{\bibfnamefont{B.}~\bibnamefont{Qi}},
  \bibinfo{author}{\bibfnamefont{S.~Q.} \bibnamefont{Zhang}},
  \bibinfo{author}{\bibfnamefont{J.}~\bibnamefont{Meng}},
  \bibinfo{author}{\bibfnamefont{S.~Y.} \bibnamefont{Wang}}, \bibnamefont{and}
  \bibinfo{author}{\bibfnamefont{S.}~\bibnamefont{Frauendorf}},
  \bibinfo{journal}{Phys. Lett. B} \textbf{\bibinfo{volume}{675}},
  \bibinfo{pages}{175} (\bibinfo{year}{2009}{\natexlab{a}}).

\bibitem[{\citenamefont{Dimitrov et~al.}(2000)\citenamefont{Dimitrov,
  Frauendorf, and D\"onau}}]{Dimitrov2000TAC}
\bibinfo{author}{\bibfnamefont{V.~I.} \bibnamefont{Dimitrov}},
  \bibinfo{author}{\bibfnamefont{S.}~\bibnamefont{Frauendorf}},
  \bibnamefont{and} \bibinfo{author}{\bibfnamefont{F.}~\bibnamefont{D\"onau}},
  \bibinfo{journal}{Phys. Rev. Lett.} \textbf{\bibinfo{volume}{84}},
  \bibinfo{pages}{5732} (\bibinfo{year}{2000}).

\bibitem[{\citenamefont{Madokoro et~al.}(2000)\citenamefont{Madokoro, Meng,
  Matsuzaki, and Yamaji}}]{Madokoro2000}
\bibinfo{author}{\bibfnamefont{H.}~\bibnamefont{Madokoro}},
  \bibinfo{author}{\bibfnamefont{J.}~\bibnamefont{Meng}},
  \bibinfo{author}{\bibfnamefont{M.}~\bibnamefont{Matsuzaki}},
  \bibnamefont{and} \bibinfo{author}{\bibfnamefont{S.}~\bibnamefont{Yamaji}},
  \bibinfo{journal}{Phys. Rev. C} \textbf{\bibinfo{volume}{62}},
  \bibinfo{pages}{061301(R)} (\bibinfo{year}{2000}).

\bibitem[{\citenamefont{Olbratowski et~al.}(2004)\citenamefont{Olbratowski,
  Dobaczewski, Dudek, and P{\l}{\'{o}}ciennik}}]{SkyrmeTAC1}
\bibinfo{author}{\bibfnamefont{P.}~\bibnamefont{Olbratowski}},
  \bibinfo{author}{\bibfnamefont{J.}~\bibnamefont{Dobaczewski}},
  \bibinfo{author}{\bibfnamefont{J.}~\bibnamefont{Dudek}}, \bibnamefont{and}
  \bibinfo{author}{\bibfnamefont{W.}~\bibnamefont{P{\l}{\'{o}}ciennik}},
  \bibinfo{journal}{Phys. Rev. Lett.} \textbf{\bibinfo{volume}{93}},
  \bibinfo{pages}{052501} (\bibinfo{year}{2004}).

\bibitem[{\citenamefont{Olbratowski et~al.}(2006)\citenamefont{Olbratowski,
  Dobaczewski, and Dudek}}]{SkyrmeTAC2}
\bibinfo{author}{\bibfnamefont{P.}~\bibnamefont{Olbratowski}},
  \bibinfo{author}{\bibfnamefont{J.}~\bibnamefont{Dobaczewski}},
  \bibnamefont{and} \bibinfo{author}{\bibfnamefont{J.}~\bibnamefont{Dudek}},
  \bibinfo{journal}{Phys. Rev. C} \textbf{\bibinfo{volume}{73}},
  \bibinfo{pages}{054308} (\bibinfo{year}{2006}).

\bibitem[{\citenamefont{Zhao}(2017)}]{ZPW2017}
\bibinfo{author}{\bibfnamefont{P.~W.} \bibnamefont{Zhao}},
  \bibinfo{journal}{Phys. Lett. B} \textbf{\bibinfo{volume}{773}},
  \bibinfo{pages}{1} (\bibinfo{year}{2017}).

\bibitem[{\citenamefont{Almehed et~al.}(2011)\citenamefont{Almehed, D\"onau,
  and Frauendorf}}]{RPA2011}
\bibinfo{author}{\bibfnamefont{D.}~\bibnamefont{Almehed}},
  \bibinfo{author}{\bibfnamefont{F.}~\bibnamefont{D\"onau}}, \bibnamefont{and}
  \bibinfo{author}{\bibfnamefont{S.}~\bibnamefont{Frauendorf}},
  \bibinfo{journal}{Phys. Rev. C} \textbf{\bibinfo{volume}{83}},
  \bibinfo{pages}{054308} (\bibinfo{year}{2011}).

\bibitem[{\citenamefont{Chen et~al.}(2013)\citenamefont{Chen, Zhang, Zhao,
  Jolos, and Meng}}]{CQB2013}
\bibinfo{author}{\bibfnamefont{Q.~B.} \bibnamefont{Chen}},
  \bibinfo{author}{\bibfnamefont{S.~Q.} \bibnamefont{Zhang}},
  \bibinfo{author}{\bibfnamefont{P.~W.} \bibnamefont{Zhao}},
  \bibinfo{author}{\bibfnamefont{R.~V.} \bibnamefont{Jolos}}, \bibnamefont{and}
  \bibinfo{author}{\bibfnamefont{J.}~\bibnamefont{Meng}},
  \bibinfo{journal}{Phys. Rev. C} \textbf{\bibinfo{volume}{87}},
  \bibinfo{pages}{024314} (\bibinfo{year}{2013}).

\bibitem[{\citenamefont{Chen et~al.}(2016)\citenamefont{Chen, Zhang, Zhao,
  Jolos, and Meng}}]{CQB2016}
\bibinfo{author}{\bibfnamefont{Q.~B.} \bibnamefont{Chen}},
  \bibinfo{author}{\bibfnamefont{S.~Q.} \bibnamefont{Zhang}},
  \bibinfo{author}{\bibfnamefont{P.~W.} \bibnamefont{Zhao}},
  \bibinfo{author}{\bibfnamefont{R.~V.} \bibnamefont{Jolos}}, \bibnamefont{and}
  \bibinfo{author}{\bibfnamefont{J.}~\bibnamefont{Meng}},
  \bibinfo{journal}{Phys. Rev. C} \textbf{\bibinfo{volume}{94}},
  \bibinfo{pages}{044301} (\bibinfo{year}{2016}).

\bibitem[{\citenamefont{Tonev et~al.}(2007)\citenamefont{Tonev, de~Angelis,
  Brant, Frauendorf, Petkov, Dewald, D\"onau, Balabanski, Zhong, Pejovic
  et~al.}}]{Tonev2007}
\bibinfo{author}{\bibfnamefont{D.}~\bibnamefont{Tonev}},
  \bibinfo{author}{\bibfnamefont{G.}~\bibnamefont{de~Angelis}},
  \bibinfo{author}{\bibfnamefont{S.}~\bibnamefont{Brant}},
  \bibinfo{author}{\bibfnamefont{S.}~\bibnamefont{Frauendorf}},
  \bibinfo{author}{\bibfnamefont{P.}~\bibnamefont{Petkov}},
  \bibinfo{author}{\bibfnamefont{A.}~\bibnamefont{Dewald}},
  \bibinfo{author}{\bibfnamefont{F.}~\bibnamefont{D\"onau}},
  \bibinfo{author}{\bibfnamefont{D.~L.} \bibnamefont{Balabanski}},
  \bibinfo{author}{\bibfnamefont{Q.}~\bibnamefont{Zhong}},
  \bibinfo{author}{\bibfnamefont{P.}~\bibnamefont{Pejovic}},
  \bibnamefont{et~al.}, \bibinfo{journal}{Phys. Rev. C}
  \textbf{\bibinfo{volume}{76}}, \bibinfo{pages}{044313}
  (\bibinfo{year}{2007}).

\bibitem[{\citenamefont{Brant et~al.}(2008)\citenamefont{Brant, Tonev,
  de~Angelis, and Ventura}}]{Brant2008}
\bibinfo{author}{\bibfnamefont{S.}~\bibnamefont{Brant}},
  \bibinfo{author}{\bibfnamefont{D.}~\bibnamefont{Tonev}},
  \bibinfo{author}{\bibfnamefont{G.}~\bibnamefont{de~Angelis}},
  \bibnamefont{and} \bibinfo{author}{\bibfnamefont{A.}~\bibnamefont{Ventura}},
  \bibinfo{journal}{Phys. Rev. C} \textbf{\bibinfo{volume}{78}},
  \bibinfo{pages}{034301} (\bibinfo{year}{2008}).

\bibitem[{\citenamefont{Raduta et~al.}(2016)\citenamefont{Raduta, Raduta, and
  Petrache}}]{CoherentJPG2016}
\bibinfo{author}{\bibfnamefont{A.~A.} \bibnamefont{Raduta}},
  \bibinfo{author}{\bibfnamefont{A.~H.} \bibnamefont{Raduta}},
  \bibnamefont{and} \bibinfo{author}{\bibfnamefont{C.~M.}
  \bibnamefont{Petrache}}, \bibinfo{journal}{J. Phys. G: Nucl. Part. Phys.}
  \textbf{\bibinfo{volume}{43}}, \bibinfo{pages}{095107}
  (\bibinfo{year}{2016}).

\bibitem[{\citenamefont{Hara and Sun}(1995)}]{Hara1995}
\bibinfo{author}{\bibfnamefont{K.}~\bibnamefont{Hara}} \bibnamefont{and}
  \bibinfo{author}{\bibfnamefont{Y.}~\bibnamefont{Sun}}, \bibinfo{journal}{Int.
  J. Mod. Phys. E} \textbf{\bibinfo{volume}{4}}, \bibinfo{pages}{637}
  (\bibinfo{year}{1995}).

\bibitem[{\citenamefont{Sabbey et~al.}(2007)\citenamefont{Sabbey, Bender,
  Bertsch, and Heenen}}]{Sabbey2007}
\bibinfo{author}{\bibfnamefont{B.}~\bibnamefont{Sabbey}},
  \bibinfo{author}{\bibfnamefont{M.}~\bibnamefont{Bender}},
  \bibinfo{author}{\bibfnamefont{G.~F.} \bibnamefont{Bertsch}},
  \bibnamefont{and} \bibinfo{author}{\bibfnamefont{P.-H.}
  \bibnamefont{Heenen}}, \bibinfo{journal}{Phys. Rev. C}
  \textbf{\bibinfo{volume}{75}}, \bibinfo{pages}{044305}
  (\bibinfo{year}{2007}).

\bibitem[{\citenamefont{Rodr\'{\i}guez and Egido}(2010)}]{Rodriguez2010}
\bibinfo{author}{\bibfnamefont{T.~R.} \bibnamefont{Rodr\'{\i}guez}}
  \bibnamefont{and} \bibinfo{author}{\bibfnamefont{J.~L.} \bibnamefont{Egido}},
  \bibinfo{journal}{Phys. Rev. C} \textbf{\bibinfo{volume}{81}},
  \bibinfo{pages}{064323} (\bibinfo{year}{2010}).

\bibitem[{\citenamefont{Rodr\'{\i}guez
  et~al.}(2015)\citenamefont{Rodr\'{\i}guez, Arzhanov, and
  Mart\'{\i}nez-Pinedo}}]{Rodriguez2015}
\bibinfo{author}{\bibfnamefont{T.~R.} \bibnamefont{Rodr\'{\i}guez}},
  \bibinfo{author}{\bibfnamefont{A.}~\bibnamefont{Arzhanov}}, \bibnamefont{and}
  \bibinfo{author}{\bibfnamefont{G.}~\bibnamefont{Mart\'{\i}nez-Pinedo}},
  \bibinfo{journal}{Phys. Rev. C} \textbf{\bibinfo{volume}{91}},
  \bibinfo{pages}{044315} (\bibinfo{year}{2015}).

\bibitem[{\citenamefont{Nik\ifmmode \check{s}\else
  \v{s}\fi{}i\ifmmode~\acute{c}\else \'{c}\fi{}
  et~al.}(2007)\citenamefont{Nik\ifmmode \check{s}\else
  \v{s}\fi{}i\ifmmode~\acute{c}\else \'{c}\fi{}, Vretenar, Lalazissis, and
  Ring}}]{Niksic2007}
\bibinfo{author}{\bibfnamefont{T.}~\bibnamefont{Nik\ifmmode \check{s}\else
  \v{s}\fi{}i\ifmmode~\acute{c}\else \'{c}\fi{}}},
  \bibinfo{author}{\bibfnamefont{D.}~\bibnamefont{Vretenar}},
  \bibinfo{author}{\bibfnamefont{G.~A.} \bibnamefont{Lalazissis}},
  \bibnamefont{and} \bibinfo{author}{\bibfnamefont{P.}~\bibnamefont{Ring}},
  \bibinfo{journal}{Phys. Rev. Lett.} \textbf{\bibinfo{volume}{99}},
  \bibinfo{pages}{092502} (\bibinfo{year}{2007}).

\bibitem[{\citenamefont{Yao et~al.}(2010)\citenamefont{Yao, Meng, Ring, and
  Vretenar}}]{YJM2010}
\bibinfo{author}{\bibfnamefont{J.~M.} \bibnamefont{Yao}},
  \bibinfo{author}{\bibfnamefont{J.}~\bibnamefont{Meng}},
  \bibinfo{author}{\bibfnamefont{P.}~\bibnamefont{Ring}}, \bibnamefont{and}
  \bibinfo{author}{\bibfnamefont{D.}~\bibnamefont{Vretenar}},
  \bibinfo{journal}{Phys. Rev. C} \textbf{\bibinfo{volume}{81}},
  \bibinfo{pages}{044311} (\bibinfo{year}{2010}).

\bibitem[{\citenamefont{Sun}(2016)}]{Sunyang2016}
\bibinfo{author}{\bibfnamefont{Y.}~\bibnamefont{Sun}}, \bibinfo{journal}{Phys.
  Scr.} \textbf{\bibinfo{volume}{91}}, \bibinfo{pages}{043005}
  (\bibinfo{year}{2016}).

\bibitem[{\citenamefont{Bhat et~al.}(2012)\citenamefont{Bhat, Sheikh, and
  Palit}}]{Bhat2012}
\bibinfo{author}{\bibfnamefont{G.~H.} \bibnamefont{Bhat}},
  \bibinfo{author}{\bibfnamefont{J.~A.} \bibnamefont{Sheikh}},
  \bibnamefont{and} \bibinfo{author}{\bibfnamefont{R.}~\bibnamefont{Palit}},
  \bibinfo{journal}{Phys. Lett. B} \textbf{\bibinfo{volume}{707}},
  \bibinfo{pages}{250} (\bibinfo{year}{2012}).

\bibitem[{\citenamefont{Bhat et~al.}(2014)\citenamefont{Bhat, Ali, Sheikh, and
  Palit}}]{Bhat2014}
\bibinfo{author}{\bibfnamefont{G.~H.} \bibnamefont{Bhat}},
  \bibinfo{author}{\bibfnamefont{R.~N.} \bibnamefont{Ali}},
  \bibinfo{author}{\bibfnamefont{J.~A.} \bibnamefont{Sheikh}},
  \bibnamefont{and} \bibinfo{author}{\bibfnamefont{R.}~\bibnamefont{Palit}},
  \bibinfo{journal}{Nucl. Phys. A} \textbf{\bibinfo{volume}{922}},
  \bibinfo{pages}{150} (\bibinfo{year}{2014}).

\bibitem[{\citenamefont{Chen et~al.}(2017)\citenamefont{Chen, Chen, Luo, Meng,
  and Zhang}}]{CFQ2017}
\bibinfo{author}{\bibfnamefont{F.~Q.} \bibnamefont{Chen}},
  \bibinfo{author}{\bibfnamefont{Q.~B.} \bibnamefont{Chen}},
  \bibinfo{author}{\bibfnamefont{Y.~A.} \bibnamefont{Luo}},
  \bibinfo{author}{\bibfnamefont{J.}~\bibnamefont{Meng}}, \bibnamefont{and}
  \bibinfo{author}{\bibfnamefont{S.~Q.} \bibnamefont{Zhang}},
  \bibinfo{journal}{Phys. Rev. C} \textbf{\bibinfo{volume}{96}},
  \bibinfo{pages}{051303} (\bibinfo{year}{2017}).

\bibitem[{\citenamefont{Shimada et~al.}(2018)\citenamefont{Shimada, Fujioka,
  Tagami, and Shimizu}}]{Shimizu2018}
\bibinfo{author}{\bibfnamefont{M.}~\bibnamefont{Shimada}},
  \bibinfo{author}{\bibfnamefont{Y.}~\bibnamefont{Fujioka}},
  \bibinfo{author}{\bibfnamefont{S.}~\bibnamefont{Tagami}}, \bibnamefont{and}
  \bibinfo{author}{\bibfnamefont{Y.~R.} \bibnamefont{Shimizu}},
  \bibinfo{journal}{Phys. Rev. C} \textbf{\bibinfo{volume}{97}},
  \bibinfo{pages}{024319} (\bibinfo{year}{2018}).

\bibitem[{\citenamefont{Ring and Schuck}(1980)}]{ManyBody}
\bibinfo{author}{\bibfnamefont{P.}~\bibnamefont{Ring}} \bibnamefont{and}
  \bibinfo{author}{\bibfnamefont{P.}~\bibnamefont{Schuck}},
  \emph{\bibinfo{title}{The Nuclear Many Body Problem}}
  (\bibinfo{publisher}{Springer-Verlag, Berlin}, \bibinfo{year}{1980}).

\bibitem[{\citenamefont{Sala et~al.}(1991)\citenamefont{Sala, Blasi, Bianco,
  Mazzoleni, Reinhardt, Schiffer, Schmittgen, Siems, and Brentano}}]{Cs130NPA}
\bibinfo{author}{\bibfnamefont{P.}~\bibnamefont{Sala}},
  \bibinfo{author}{\bibfnamefont{N.}~\bibnamefont{Blasi}},
  \bibinfo{author}{\bibfnamefont{G.~L.} \bibnamefont{Bianco}},
  \bibinfo{author}{\bibfnamefont{A.}~\bibnamefont{Mazzoleni}},
  \bibinfo{author}{\bibfnamefont{R.}~\bibnamefont{Reinhardt}},
  \bibinfo{author}{\bibfnamefont{K.}~\bibnamefont{Schiffer}},
  \bibinfo{author}{\bibfnamefont{K.}~\bibnamefont{Schmittgen}},
  \bibinfo{author}{\bibfnamefont{G.}~\bibnamefont{Siems}}, \bibnamefont{and}
  \bibinfo{author}{\bibfnamefont{P.~V.} \bibnamefont{Brentano}},
  \bibinfo{journal}{Nucl. Phys. A} \textbf{\bibinfo{volume}{531}},
  \bibinfo{pages}{383 } (\bibinfo{year}{1991}).

\bibitem[{\citenamefont{Kumar et~al.}(2001)\citenamefont{Kumar, Mehta, Singh,
  Kaur, G{\"o}rgen, Chmel, Singh, and Murlithar}}]{Cs130EPJA}
\bibinfo{author}{\bibfnamefont{R.}~\bibnamefont{Kumar}},
  \bibinfo{author}{\bibfnamefont{D.}~\bibnamefont{Mehta}},
  \bibinfo{author}{\bibfnamefont{N.}~\bibnamefont{Singh}},
  \bibinfo{author}{\bibfnamefont{H.}~\bibnamefont{Kaur}},
  \bibinfo{author}{\bibfnamefont{A.}~\bibnamefont{G{\"o}rgen}},
  \bibinfo{author}{\bibfnamefont{S.}~\bibnamefont{Chmel}},
  \bibinfo{author}{\bibfnamefont{R.}~\bibnamefont{Singh}}, \bibnamefont{and}
  \bibinfo{author}{\bibfnamefont{S.}~\bibnamefont{Murlithar}},
  \bibinfo{journal}{Eur. Phys. J. A} \textbf{\bibinfo{volume}{11}},
  \bibinfo{pages}{5} (\bibinfo{year}{2001}).

\bibitem[{\citenamefont{Simons et~al.}(2005)\citenamefont{Simons, Joshi,
  Jenkins, Raddon, Wadsworth, Fossan, Koike, Vaman, Starosta, Paul
  et~al.}}]{Cs130transitions}
\bibinfo{author}{\bibfnamefont{A.~J.} \bibnamefont{Simons}},
  \bibinfo{author}{\bibfnamefont{P.}~\bibnamefont{Joshi}},
  \bibinfo{author}{\bibfnamefont{D.~G.} \bibnamefont{Jenkins}},
  \bibinfo{author}{\bibfnamefont{P.~M.} \bibnamefont{Raddon}},
  \bibinfo{author}{\bibfnamefont{R.}~\bibnamefont{Wadsworth}},
  \bibinfo{author}{\bibfnamefont{D.~B.} \bibnamefont{Fossan}},
  \bibinfo{author}{\bibfnamefont{T.}~\bibnamefont{Koike}},
  \bibinfo{author}{\bibfnamefont{C.}~\bibnamefont{Vaman}},
  \bibinfo{author}{\bibfnamefont{K.}~\bibnamefont{Starosta}},
  \bibinfo{author}{\bibfnamefont{E.~S.} \bibnamefont{Paul}},
  \bibnamefont{et~al.}, \bibinfo{journal}{J. Phys. G: Nucl. Part. Phys.}
  \textbf{\bibinfo{volume}{31}}, \bibinfo{pages}{541} (\bibinfo{year}{2005}).

\bibitem[{\citenamefont{Gao et~al.}(2006)\citenamefont{Gao, Chen, and
  Sun}}]{GZC2006}
\bibinfo{author}{\bibfnamefont{Z.~C.} \bibnamefont{Gao}},
  \bibinfo{author}{\bibfnamefont{Y.}~\bibnamefont{Chen}}, \bibnamefont{and}
  \bibinfo{author}{\bibfnamefont{Y.}~\bibnamefont{Sun}},
  \bibinfo{journal}{Phys. Lett. B} \textbf{\bibinfo{volume}{634}},
  \bibinfo{pages}{195} (\bibinfo{year}{2006}).

\bibitem[{\citenamefont{Sheikh et~al.}(2008)\citenamefont{Sheikh, Bhat, Sun,
  Vakil, and Palit}}]{Sheikh2008}
\bibinfo{author}{\bibfnamefont{J.~A.} \bibnamefont{Sheikh}},
  \bibinfo{author}{\bibfnamefont{G.~H.} \bibnamefont{Bhat}},
  \bibinfo{author}{\bibfnamefont{Y.}~\bibnamefont{Sun}},
  \bibinfo{author}{\bibfnamefont{G.~B.} \bibnamefont{Vakil}}, \bibnamefont{and}
  \bibinfo{author}{\bibfnamefont{R.}~\bibnamefont{Palit}},
  \bibinfo{journal}{Phys. Rev. C} \textbf{\bibinfo{volume}{77}},
  \bibinfo{pages}{034313} (\bibinfo{year}{2008}).

\bibitem[{\citenamefont{Wang et~al.}(2007)\citenamefont{Wang, Zhang, Qi, and
  Meng}}]{WSY2007}
\bibinfo{author}{\bibfnamefont{S.~Y.} \bibnamefont{Wang}},
  \bibinfo{author}{\bibfnamefont{S.~Q.} \bibnamefont{Zhang}},
  \bibinfo{author}{\bibfnamefont{B.}~\bibnamefont{Qi}}, \bibnamefont{and}
  \bibinfo{author}{\bibfnamefont{J.}~\bibnamefont{Meng}},
  \bibinfo{journal}{Chin. Phys. Lett.} \textbf{\bibinfo{volume}{24}},
  \bibinfo{pages}{664} (\bibinfo{year}{2007}).

\bibitem[{\citenamefont{Frauendorf and Meng}(1996)}]{ZPA1996}
\bibinfo{author}{\bibfnamefont{S.}~\bibnamefont{Frauendorf}} \bibnamefont{and}
  \bibinfo{author}{\bibfnamefont{J.}~\bibnamefont{Meng}}, \bibinfo{journal}{Z.
  Phys. A} \textbf{\bibinfo{volume}{356}}, \bibinfo{pages}{263}
  (\bibinfo{year}{1996}).

\bibitem[{\citenamefont{Wang et~al.}(2009)\citenamefont{Wang, Wu, Zhu, Li, Hao,
  Zheng, He, Wang, Li, Liu et~al.}}]{Cs130-ZhuLH}
\bibinfo{author}{\bibfnamefont{L.~L.} \bibnamefont{Wang}},
  \bibinfo{author}{\bibfnamefont{X.~G.} \bibnamefont{Wu}},
  \bibinfo{author}{\bibfnamefont{L.~H.} \bibnamefont{Zhu}},
  \bibinfo{author}{\bibfnamefont{G.~S.} \bibnamefont{Li}},
  \bibinfo{author}{\bibfnamefont{X.}~\bibnamefont{Hao}},
  \bibinfo{author}{\bibfnamefont{Y.}~\bibnamefont{Zheng}},
  \bibinfo{author}{\bibfnamefont{C.~Y.} \bibnamefont{He}},
  \bibinfo{author}{\bibfnamefont{L.}~\bibnamefont{Wang}},
  \bibinfo{author}{\bibfnamefont{X.~Q.} \bibnamefont{Li}},
  \bibinfo{author}{\bibfnamefont{Y.}~\bibnamefont{Liu}}, \bibnamefont{et~al.},
  \bibinfo{journal}{Chin. Phys. C} \textbf{\bibinfo{volume}{33}},
  \bibinfo{pages}{173} (\bibinfo{year}{2009}).

\bibitem[{\citenamefont{Qi et~al.}(2009{\natexlab{b}})\citenamefont{Qi, Zhang,
  Wang, Yao, and Meng}}]{QB2009PRC}
\bibinfo{author}{\bibfnamefont{B.}~\bibnamefont{Qi}},
  \bibinfo{author}{\bibfnamefont{S.~Q.} \bibnamefont{Zhang}},
  \bibinfo{author}{\bibfnamefont{S.~Y.} \bibnamefont{Wang}},
  \bibinfo{author}{\bibfnamefont{J.~M.} \bibnamefont{Yao}}, \bibnamefont{and}
  \bibinfo{author}{\bibfnamefont{J.}~\bibnamefont{Meng}},
  \bibinfo{journal}{Phys. Rev. C} \textbf{\bibinfo{volume}{79}},
  \bibinfo{pages}{041302(R)} (\bibinfo{year}{2009}{\natexlab{b}}).

\end{thebibliography}

\end{CJK}

\end{document}